\documentclass[a4paper]{jpconf}

\usepackage{url}
\newbox\shell
\newcommand{\dia}[2]{\setbox\shell=\hbox{\begin{picture}(180,220)(-90,-110)#1
\put(-90,-95){\makebox(180,220)[b]{#2}}\end{picture}}\dimen0=\ht
\shell\multiply\dimen0by7\divide\dimen0by16\raise-\dimen0\box\shell\hfill}
\newcommand{\feyndia}{
\put(-200,-50){\line(1,0){100}}
\put(100,-50){\line(-1,1){100}}
\put(100,-50){\line(1,0){100}}
\put(-100,-50){\line(1,1){100}}
\put(-40,60){\mbox{magnet}}
\put(50,10){\mbox{$e$}}
\put(-60,10){\mbox{$e$}}
\put(-150,-80){\mbox{$e$}}
\put(150,-80){\mbox{$e$}}
\put(-40,-80){\mbox{photon}}
\linethickness{0.5mm}
\put(-100,-50){\line(1,0){200}}}

\begin{document}

\title{Richard Feynman's talent for finding things out}

\author{David Broadhurst}

\address{Open University, Milton Keynes MK76AA, UK}

\ead{David.Broadhurst@open.ac.uk}

\begin{abstract}
This article is a summary of a talk about Richard Feynman, given at a conference {\em Polymaths across the Eras}
organized in November 2023 by the St Cross Centre for the History and Philosophy of Physics (HAPP) in Oxford.
It describes Feynman as an unconventional polymath, 
primarily concerned with his own understanding of nature, having little regard for previous authorities.
His curiosity, respect for experiment, ability to calculate and notable ingenuity led to important developments 
in quantum mechanics, quantum electrodynamics, and strong interactions.
In particle physics, Feynman diagrams have been a mainstay of calculation for the last 75 years. 
His relish for the spoken word led to renown as an educator and raconteur.
His perceptions were influential in a wide range of other scientific fields, including weak interactions, gravity, 
superconductivity, biology, nanotechnology, algorithmic computation and quantum computers. 
In all of this, he held to the idea that nature has an intrinsic simplicity and beauty that permit us to find things out,
if only we will try hard enough.
\end{abstract}

\section{Introduction}

Richard Phillips Feynman (1918--1988) might have dismissed the idea that he was a polymath, 
in the dictionary sense -- {\em a person of wide learning} -- 
or even in an informal sense of being good at lots of different things.

What distinguished him among 20th century scientists was his determination to find things out, 
on his own terms and for his own satisfaction, 
sometimes with scant regard for previous authorities or research by contemporaries. 
He combined this approach with a respect for the primacy of experiment, a formidable ability to calculate and considerable ingenuity.
Here are tributes from physicists of high repute, taken from the book {\em Most of the Good Stuff}~\cite{MGS}.
\begin{itemize}
\item Hans Bethe (1906--2005): Feynman is like a magician who ``does things that nobody else 
could ever do and that seem completely unexpected."
\item Julian Schwinger (1918--1994): ``an honest man, the outstanding 
intuitionist of our age and a prime example of what may lie in 
store for anyone who dares to follow the beat of a different drum."
\item Freeman Dyson (1923--2020): ``the most original mind of his generation." 
{\em After} Ben Jonson on Shakespeare: ``I did love the man this side idolatry as much as any other.''
\item Murray Gell-Mann (1929--2019): He ``liked to look at each problem, 
important or unimportant, in a new way -- {\em turning it around}.
But with Dick, turning things around and being different became a passion."
\end{itemize}

My Stanford mentors Jim Bjorken and Sid Drell held him in high regard. 
From their book, {\em Relativistic Quantum Fields}~\cite{BjD}, 
I had learnt about Feynman diagrams.

As an educator and a raconteur Feynman enjoyed explaining things that he had found out for himself, 
often speaking and writing colloquially, as here: 
\begin{itemize}
\item {\em The Feynman Lectures on Physics}~\cite{FLP}, given at Caltech 1961--1963, 
\item {\em The Character of Physical Law}~\cite{CPL}, Messenger lectures at Cornell 1964,
\item {\em The world from another point of view}, Yorkshire TV 
\url{http://youtu.be/GNhlNSLQAFE} 1973,
\item {\em The pleasure of finding things out}, BBC TV 
\url{http://vimeo.com/340695809} 1981, 
\item {\em QED: The Strange Theory of Light and Matter}~\cite{QED}, 1983, 
\item {\em Surely You're Joking, Mr Feynman!}~\cite{SYJ}, 1985, 
\item {\em What Do You Care What Other People Think?}~\cite{WDY}, 1988,
\item {\em An Outsider's Inside View of the Challenger Inquiry}~\cite{SSC}, 1988.
\end{itemize}

\section{Limitations and appreciation} 

Unlike Bethe, his methods were sometimes unclear to others. 
Unlike Schwinger, he did not nourish a large cohort of eminent PhD students. 
Unlike Dyson, he neglected deductive reasoning, preferring predictive answers. 
Unlike Oppenheimer and Gell-Mann, he had little interest in the humanities. 
His grades in English and history were abysmal; those in maths and physics were amazing.
He preferred the spoken to the written word.
Robert Leighton and Matthew Sands helped him to write up his {\em Lectures on Physics}. 
Feynman's preface takes the ``pessimistic'' view that they were unsuccessful for a majority of undergraduates. 
He encouraged a ``reputation for irresponsibility'', so as to be free from administration. 
After 1973, he made few new contributions to the standard model of particle physics. 
He had little liking for speculations on supersymmetry or string theory. 
His relationships with other notable physicists were sometimes difficult.

To my mind, such limitations are the reverse of a coin upon whose obverse he stamped
a vivid and indelible portrait of a mind constantly seeking to refine his
own understanding of nature. He strove to find alternative ways
of arriving at quantitative predictions. He explored a variety of calculational methods
whose diversity might inform future work on unsolved problems.

He was {\em sui generis}.

\section{Overview of Feynman's work}

The title of James Gleick's 1992 biography~\cite{JGG} of Feynman is {\em Genius}. 
Here, I should like to invoke Samuel Johnson's aphorism: 
``the {\em true} genius is a mind of large general powers, accidentally determined to some particular direction.''

I shall focus  on what I regard as Feynman's most significant contributions to my own subject, particle physics:\\
1) quantum theory as sums over histories,\\ 
2) Feynman diagrams for ever,\\ 
3) the parton model of protons and neutrons.

I shall mention work that shows him to be a seeker-out in various other fields:\\
4) weak interactions,\\ 
5) gravity and quantum theory,\\ 
6) superconductivity and superfluidity,\\ 
7) biology and nanotechnology,\\ 
8) algorithmic computation for the Manhattan project,\\ 
9) the possibility of quantum computers.

\subsection{Sums over histories}

Feynman liked to boil down quantum theory to the following prescription~\cite{ST1,ST2,FNP,QPI}.
To calculate the probability that an initial state {\em A} results in a final state {\em B},
imagine all the ways that this {\em might} happen;
find a rule that assigns a complex {\em number} to each;
if you cannot distinguish between histories, take the modulus {\em squared} of the {\em sum} of these numbers; 
if you can detect a difference, you should take a sum of squares.

After working on his PhD, with advice from John Wheeler (1911--2008), 
Feynman came to believe that this was a key to all quantum phenomena 
and sought to use it as much as possible, in some cases with only limited success.

When there are only two histories to consider, the maths becomes much easier.

\subsubsection*{Anecdote 1} 

I arrived at Oxford in 1965 at the same time that 
Volume III of the {\em Feynman Lectures} arrived at Blackwell's bookshop.
I was an (almost) classical virgin. In those days, 
no quantum physics was needed for the Oxford scholarship examination; 
quantum theory was considered X-rated material, unsuitable for schoolboys. 
I agreed to a novel pedagogical experiment devised by Paul Bamberg. 
I would study no other book on quantum mechanics until I had mastered
the first 12 chapters of Volume III. Feynman gave me a perspective that abides.
Like many other university teachers, I have returned to these lectures, 
for novel and illuminating ways of reaching subsequent generations of physics students.
 
\subsection{Feynman diagrams}
 
Sin-Itiro Tomonaga (1906--1979), Schwinger and Feynman shared the physics Nobel prize of 1965,
{\em for their fundamental work in quantum electrodynamics, 
with deep-ploughing consequences for the physics of elementary particles.} 
Tomonaga's work in war-torn Japan was not widely known when Schwinger achieved his first success in 1947.
Feynman's approach appeared, at first sight, to be rather different from Schwinger's~\cite{JSP}.
In a remarkable feat of mathematical synthesis, Dyson digested, reconciled 
and extended their methods of analysis~\cite{FJD}.

The object of the exercise was to find successive corrections~\cite{BBF} to older predictions, 
by expanding in the fine structure constant, $\alpha\approx1/137$, 
formed from Planck's constant, the speed of light and the electron's charge. 
For example, Paul Dirac (1902--1984) had obtained a magnetic moment for the electron 
of exactly 1 Bohr magneton. Schwinger computed a radiative correction, 
obtaining $1+\alpha/(2\pi)\approx1.00116.$ 
There were three main parts to such challenges:\\ 
A) derive, from clearly stated principles, the rules for such calculations;\\ 
B) devise ways of getting rid of infinities produced by such rules;\\ 
C) evaluate the demanding integrals that give the required answers.

Feynman skipped step A and guessed his rules, depicted by diagrams.

I found, in {\tt arXiv}, 7646 articles with {\em Feynman}, in titles, author lists, or abstracts,
very often followed by {\em diagram}.
He would be happy that astrophysicist Joan Feynman (1927--2020) appears.
Without his sister's cheerful chiding, he would have published less.
Encouragement also came from C\'ecile Morette (1922--2017). 
With Dyson, she went to Feynman in 1948, asking about two baffling diagrams. 
Dyson told his parents that Feynman ``proceeded to sit down and in two hours, 
before our very eyes, obtain finite and sensible answers to both problems.''

It took some effort from Dyson to convince Robert Oppenheimer (1904--1967) that Feynman diagrams
and their rules are equivalent to Schwinger's derivations of perturbative expansions in quantum electrodynamics.
Between 1949 and 1954, the number of  {\em Physical Review} articles using Feynman diagrams increased exponentially,
doubling about every two years~\cite{DTA}. 
C\'ecile Morette's article~\cite{CMD} appeared in the same volume as Feynman's~\cite{ST2}.
For the last 75 years, Feynman diagrams have been ubiquitous in the organization of perturbative expansions.

\subsubsection*{Anecdote 2}
 
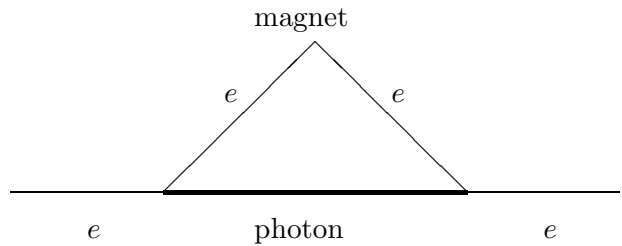
\mbox{\hspace{5cm}}\dia{\feyndia}{}\\ 
In 1968 my thesis advisor Gabriel Barton (1934--2022) invited me to calculate 
Schwinger's radiative correction, ${0.00116}$, with books closed. 
{\em Easy}, I thought. I knew the diagram and the rules for its lines and vertices. 
So I had it as a finite integral over momentum. But how to evaluate that?
I went back to Gabriel, who showed me, in a few seconds, Feynman's trick for combining lines.
I was then set up for the next 55 years of work on more difficult diagrams.

\subsection{Parton model}
 
The kernel of this model was the idea that nucleons (protons or neutrons) interact with themselves 
and with leptons (electrons or neutrinos) in a manner that may reveal simpler parts of those complex nucleons.
When I arrived at Stanford in 1971, experimental data~\cite{DIS} had accumulated to support this idea.
Bjorken and Drell were working on it, but I heard people refer to {\em Feynman's} parton model. 
Yet he seemed to write or talk about it sparingly.
He had written in 1969~\cite{VHE}:
``I am more sure of the conclusions than of any single argument which suggested them 
to me for they have an internal consistency which surprises me and exceeds 
the consistency of my deductive arguments which hinted at their existence.''
I did not see the word  {\em parton} in his written work until 1972~\cite{PHI}.  
Was Feynman still working hard on this? Who knew? 
Apparently, not even Bjorken, whose own published work was pivotal and explicit~\cite{BjS,BjP}.

\subsubsection*{Anecdote 3}

The mist began to clear when I heard Feynman talk in a discussion forum at Irvine,
at first reluctantly, then more freely, always informally. 
Roger Cashmore and I took house-mates to hear Feynman talk at Berkeley. 
``Why can't you explain things as clearly and as simply as Feynman does?'' 
asked these bright life-scientists. 
``Because he is Feynman'', I ruefully replied. 

\subsection{Weak interactions}
 
The weak interaction is responsible for the beta-decay of a neutron, 
leaving a proton, an electron, and a neutrino that may pass through the Earth. 
In 1956 it was found, to everyone's great surprise,
that weak processes distinguish left from right: their mirror images are impossible. 
Feynman and Gell-Mann worked together on the rules for weak interactions, 
sometimes using amateur radio to communicate between Brazil and California. 
There were 5 types to consider, labelled {\em A, P, S, T, V}. 
They arrived at the combination $(V-A)$~\cite{FGM}. 
So did other people. 
Here is Feynman's verdict on priority~\cite{VMA}: 
``We have a conventional theory of weak interactions invented by Marshak and Sudarshan,
published by Feynman and Gell-Mann and completed by Cabibbo.'' Feynman seemed less
bothered about priority than others of his generation.

\subsection {Quantum theory of gravitation}
 
 Feynman studied all four interactions: electromagnetic, weak, strong and gravitational,
the last being the hardest to reconcile with quantum theory. 
Being Feynman, he was one of the first to try to handle this tough challenge.  
His article on the subject, in 1963, combines modesty with ambition~\cite{QTG}. 
Unable to deal with infinities (manageable in electrodynamics) he wrote: 
``I will carry out the perturbation expansion as far as I can in every direction.''

\subsection{Superconductivity and superfluidity} 

Bellow a certain critical temperature, some solids can sustain electrical currents
with no loss of power and some liquids can flow without friction.
Feynman worked hard and long on such problems, considering them very important.
In September 1956 he give a talk~\cite{SFC}, discussing his failures and concluding: 
``the only reason that we cannot do this problem of superconductivity is that 
we haven't got enough imagination.'' A few months later, in February 1957, 
John Bardeen (1908--1991), Leon Cooper (1930--) and John Schrieffer (1931--2019)
submitted an article {\em Microscopic Theory of Superconductivity}~\cite{BCS}, 
with the leap of imagination that won them the physics Nobel prize of 1972.
I like to think that Feynman might have applauded their ingenuity.

\subsection{Biology and nanotechnology}
 
Studying plant cells under a microscope, Feynman asked: 
``How do they circulate? What pushes them around?" 
He remarked that in biology
``it was very easy to find a question that was very interesting, and that nobody knew the answer to.'' 
Seeing that microscopic organisms do intricate things, he spent a year in the laboratory of Max Delbr\"uck (1906--1981), 
studying mutations of viruses~\cite{EFK}. 
His 1959 talk {\em There's plenty of room at the bottom}~\cite{PRB} set an agenda for nanotechnolgy: 
can we manipulate matter on an almost atomic scale to make new materials?

\subsection{Algorithmic computation at Los Alamos}

Under the threat of atomic-bomb developments in war-time Germany,
the physics talent of North America and UK set aside pure research, for the duration. 
For Feynman, at age 23 in 1941, this was a huge dent in his burgeoning career. 
His wife Arline Greenbaum (1919--1945) died after only 3 years of marriage~\cite{OH5}.  
It was she, not he, who asked: ``what do {\em you} care what {\em other} people think?''
Richard brought his considerable powers of concentration and computation
to the Manhattan project, as recounted in his 1975 talk {\em Los Alamos from below}~\cite{LAB}. 
He developed algorithms for vital calculations, with error correction. 
He assembled electromagnetic computers and became an expert safe-breaker.

\subsection{Quantum computation}
 
It was thus natural that Feynman was one of the first people to enquire about modification
of classical computers by the uncertainties of quantum mechanics~\cite{SPC}.
Nowadays, many people hold out hopes for exploiting quantum theory,
to achieve enormous speed-ups, in comparison with classical machines. 
They regard Feynman and Yuri Manin (1937--2023) as early pioneers. 
As before, we see Feynman's propensity to anticipate, stimulate and provoke.

\section{Feynman's delight in nature}

In an inspiring lecture on {\em The Character of Physical Law}, he said:
``In this age people are experiencing a delight, 
the tremendous delight that you get when you guess 
how nature will work in a new situation never seen before. 
From experiments and information in a certain range 
you can guess what is going to happen in a region 
where no one has ever explored before. 
It is a little different from regular exploration in that there are  
enough clues on the land discovered to guess what the  
land that has not been discovered is going to look like. 
These guesses, incidentally, are often very different from 
what you have already seen -- they take a lot of thought. 
What is it about nature that lets this happen, 
that it is possible to guess from one part 
what the rest is going to do? 
That is an unscientific question:
I do not know how to answer it, 
and therefore I am going to give an unscientific answer. 
 I think it is because nature has a simplicity
and therefore a great beauty."

\section{Conclusion}
 
Like all of us, Feynman had flaws: intellectual, social and moral. 
Several of these he would readily admit. I shall not dwell on them. 
He questioned, informed, annoyed and inspired
many other scientists, across a wide range of human enquiry, 
with a strong commitment to finding things out 
that merits him to be considered as an unconventional polymath.

\subsection*{Acknowledgments}

I thank Ian Aitchison, Joanna Ashbourn, Lance Dixon and Graham Farmelo for generous advice.

\raggedright

\end{document}